\documentclass[twocolumn,pra,aps]{revtex4}
\usepackage{graphicx}

\def\duzomniejsze{<\kern-.7mm<}
\def\duzowieksze{>\kern-.7mm>}

\def\textbf#1{{\bf #1}}
\def\beq{\begin{equation}}
\def\eeq{\end{equation}}
\def\be{\begin{equation}}
\def\ee{\end{equation}}
\def\ben{\begin{eqnarray}}
\def\een{\end{eqnarray}}
\def\beqa{\begin{eqnarray}}
\def\eeqa{\end{eqnarray}}
\def\eea{\end{array}}
\def\bea{\begin{array}}
\newcommand{\bei}{\begin{itemize}}
\newcommand{\eei}{\end{itemize}}
\newcommand{\bee}{\begin{enumerate}}
\newcommand{\eee}{\end{enumerate}}

\def\>{\rangle}
\def\<{\langle}

\begin{document}

\title{Nonclassicality of single quantum excitation of a thermal field in thermal environments}

\begin{abstract}
The nonclassicality of single photon-added thermal states in the
thermal channel is investigated by exploring the volume of the
negative part of the Wigner function. The Wigner functions become
positive when the decay time exceeds a threshold value
$\gamma{t}_c$, which only depends on the effective temperature or
mean thermal photon number of the thermal channel, but not depends
on the effective temperature of the initial thermal state. This
phenomenon is similar with the case of single quantum excitation of
classical coherent states in thermal channel. Furthermore, we
firstly demonstrate $\gamma{t}_c$ is the same for arbitrary pure or
mixed nonclassical optical fields with zero population in vacuum
state.

\end{abstract}
\author{Shang-Bin Li}

\affiliation{Shanghai research center of Amertron-global,
Zhangjiang High-Tech Park, \\
299 Lane, Bisheng Road, No. 3, Suite 202, Shanghai, 201204, P.R.
China}

\maketitle

The preparations of nonclassical non-gaussian optical fields have
attracted much attention, which have many applications in quantum
information processing \cite{Bouwmeester}. Usually, the
nonclassicality manifests itself in specific properties of quantum
statistics \cite{Kimble,Short,Dodonov,Hillery}, in which the partial
negative Wigner distribution \cite{Hillery} is indeed a good
indication of the highly nonclassical character of the optical
fields. Reconstruction of the Wigner distribution in experiments
with quantum tomography \cite{Vogel,Smithey,Welsch} have
demonstrated the appearance of the negative values, which does not
have any classical counterparts. A variety of nonclassical states
has recently been characterized by means of the negativity of their
Wigner function
\cite{Lvovsky,Zavatta01,Zavatta02,Zavatta03,Kim2005}, which is a
sufficient and not necessary condition for nonclassicality
\cite{Lvovsky02}.

The photon-added thermal states (PATSs) were introduced by Agarwal
and Tara \cite{Agarwal1992}, which do not exhibit squeezing,
sub-Poissonian statistics, and any coherence. The single
photon-added thermal state (SPATS) has been experimentally prepared
by Zavatta et al. and its nonclassical properties have been detected
by homodyne tomography technology \cite{Zavatta2007}. For the
SPATSs, their nonclassical properties have been investigated by
several authors \cite{Agarwal1992,Jones1997,Zavatta2007,Devi}.
Furthermore, Parigi et al. have experimentally investigated quantum
commutation rules by addition and subtraction of single photons to
or from a light field initially in the thermal state
\cite{Zavatta-s}. Ordinarily, the interaction between the
nonclassical optical fields and their surrounding thermal
environment may deteriorate the degree of nonclassicality. Thus, to
study the dynamical behaviors of the partial negativity of Wigner
distribution and understand how long a nonclassical field preserves
its partial negative Wigner distribution in thermal channel may be
very desirable for experimentally quantifying the variation of
nonclassicality. Here, the nonclassicality of photon-added thermal
states in the thermal channel is investigated by exploring the
partial negative Wigner distribution. The exact expression of the
time evolution Wigner function is given out and the dynamical
behavior of the volume of the negative part of the Wigner function
is analytically derived. The threshold values $\gamma{t}$ of the
decay time corresponding to the transition of the Wigner
distribution from partial negative to completely positive are
derived. For SPATSs in thermal channel, it is shown that the
threshold value of the decay time is independent of the mean thermal
photon number of initial thermal state seed. Furthermore, we firstly
demonstrate $\gamma{t}_c$ is the same for arbitrary pure or mixed
nonclassical optical fields with zero population in vacuum state,
and is given by $\gamma{t}_c=\ln(\frac{2+2n}{1+2n})$, where $n$ is
the mean thermal photon number of the thermal channel.

Let us first briefly recall the definition of the single
photon-added thermal states (SPATSs) \cite{Agarwal1992}. The SPATSs
are defined by \be
\hat{\rho}=\frac{1}{\bar{n}(\bar{n}+1)}\sum^{\infty}_{l=0}\frac{\bar{n}^l}{(1+\bar{n})^{l}}l|l\rangle\langle{l}|,
\ee where $|l\rangle$ is the Fock state and $\bar{n}$ is the mean
photon number of the thermal state seed. When the SPATS evolves in
the thermal channel, the evolution of the density matrix can be
described by \cite{Gardiner} \beqa
\frac{d\hat{\rho}}{dt}&=&\frac{\gamma(n+1)}{2}(2a\hat{\rho}{a}^{\dagger}-a^{\dagger}a\hat{\rho}-\hat{\rho}{a}^{\dagger}a)\nonumber\\
&&+\frac{\gamma{n}}{2}(2a^{\dagger}\hat{\rho}{a}-aa^{\dagger}\hat{\rho}-\hat{\rho}{a}{a}^{\dagger}),
\eeqa where $\gamma$ represents dissipative coefficient and $n$
denotes the mean thermal photon number of the thermal channel.
$a^{\dagger}$ ($a$) is the creation operator (annihilation operator)
of the optical mode. When $n=0$, the Eq.(2) reduces to the master
equation describing the photon-loss channel.

For an optical field in the state $\hat{\rho}$, its Wigner function,
the Fourier transformation of characteristics function
\cite{Barnett} of the state $\hat{\rho}$ can be derived by
\cite{Cessa,Englert} \be
W(\beta)=\frac{2}{\pi}{\mathrm{Tr}}[(\hat{O}_e-\hat{O}_o)\hat{D}(\beta)\hat{\rho}\hat{D}^{\dagger}(\beta)],\ee
where $\hat{O}_e\equiv\sum^{\infty}_{k=0}|2k\rangle\langle2k|$ and
$\hat{O}_o\equiv\sum^{\infty}_{k=0}|2k+1\rangle\langle2k+1|$ are the
even and odd parity operators respectively. In the thermal channel
described by the master Eq.(2), the time evolution Wigner function
satisfies the following Fokker-Planck equation \cite{Carmichael}
\beqa
\frac{\partial}{\partial{t}}W(q,p,t)&=&\frac{\gamma}{2}(\frac{\partial}{\partial{q}}q+\frac{\partial}{\partial{p}}p)W(q,p,t)\nonumber\\
&&+\frac{\gamma(2n+1)}{8}(\frac{\partial^2}{\partial{q}^2}+\frac{\partial^2}{\partial{p}^2})W(q,p,t).\eeqa
where $q$ and $p$ represent the real part and imaginary part of
$\beta$, respectively. The time evolution Wigner function can be
derived as following: \beqa
W(q,p,\gamma{t})&=&\exp(\gamma{t})\int^{\infty}_{-\infty}\int^{\infty}_{-\infty}W_T(x,y)\nonumber\\
&&\cdot{W}(\frac{q-\sqrt{1-e^{-\gamma{t}}}x}{\sqrt{e^{-\gamma{t}}}},\frac{p-\sqrt{1-e^{-\gamma{t}}}y}{\sqrt{e^{-\gamma{t}}}},0)dxdy\nonumber\\
\eeqa where \be
W_T(x,y)=\frac{2}{\pi(1+2n)}\exp(-\frac{2(x^2+y^2)}{1+2n})\ee is the
Wigner function of the thermal state with mean photon number $n$.
Substituting the initial Wigner function of the SPATS
\cite{Agarwal1992} \be
W^S(q,p,0)=\frac{2}{\pi}(\frac{4(1+\bar{n})(q^2+p^2)}{(1+2\bar{n})^3}-\frac{1}{(1+2\bar{n})^2})e^{-\frac{2}{1+2\bar{n}}(q^2+p^2)}\ee
into the Eq.(5), it is easy to obtain the corresponding time
evolution Wigner function as follows: \beqa
W^S(q,p,\gamma{t})&=&\frac{\kappa+8(1+\bar{n})e^{\gamma{t}}(q^2+p^2)}{\pi{\xi^3}}e^{\frac{\zeta-2e^{\gamma{t}}(q^2+p^2)}{\xi}},\nonumber\\
\xi&=&2(\bar{n}-n)+(1+2n)e^{\gamma{t}},\nonumber\\
\zeta&=&2(\bar{n}-n)\gamma{t}+(1+2n)\gamma{t}e^{\gamma{t}},\nonumber\\
\kappa&=&-8(\bar{n}-n)(1+n)+2(1+2n)^2e^{2\gamma{t}}\nonumber\\
&&+4[\bar{n}(1+2n)-(1+2n)^2]e^{\gamma{t}}. \eeqa In Fig.1, the
Wigner function of the SPATS with $\bar{n}=1$ in the thermal channel
with $n=0.5$ is plotted for three different values of decay time. It
is shown that the phase space Wigner distribution of the SPATS
exhibits partial negativity around the origin, and the region of the
negative part in phase space is a circle. The absolute value of
negative minimum of the Wigner distribution decreases as $\gamma{t}$
increases, and the thermal noise causes the disappearance of the
partial negativity of the Wigner function if the decay time exceeds
a threshold value. The ringlike wings in the distribution gradually
disappear and the distribution becomes more and more similar to the
Gaussian typical of a thermal state.

\begin{figure}
\centerline{\includegraphics[width=6.3cm]{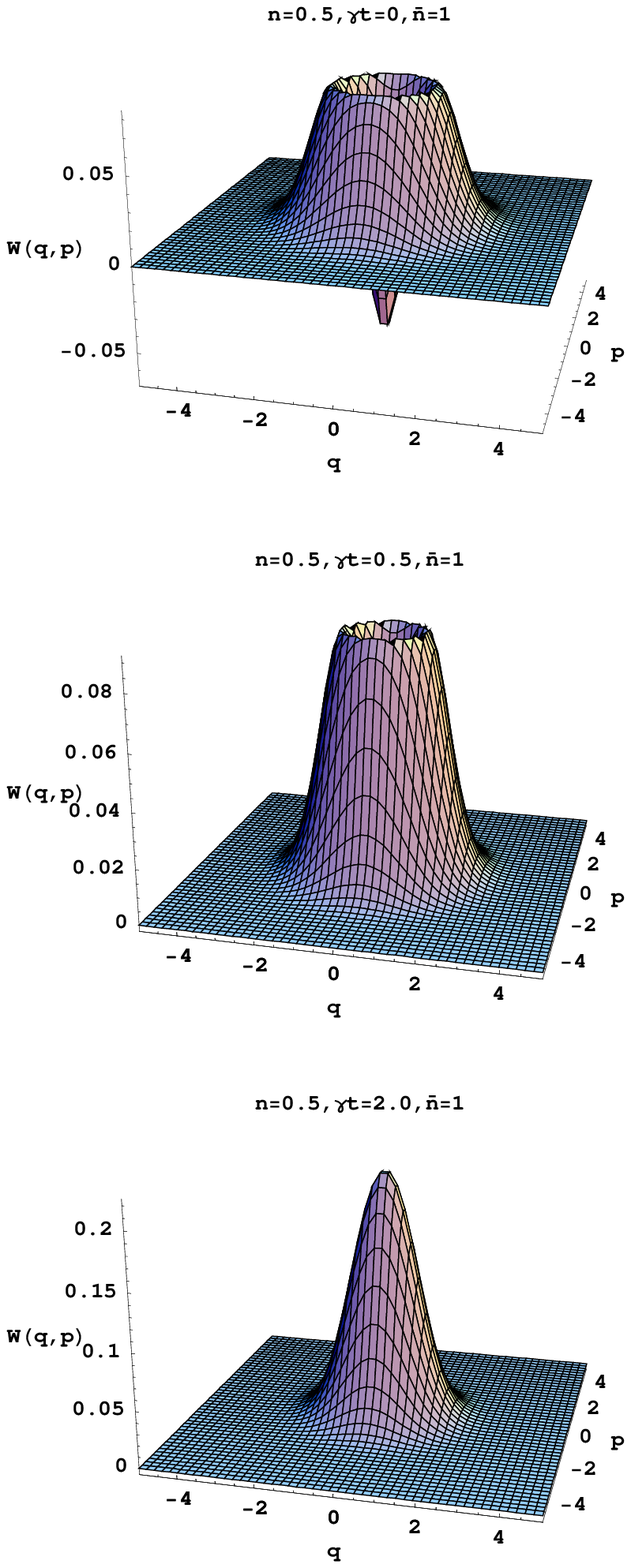}}
\caption{The Wigner functions of the SPATS with $\bar{n}=1$ in
thermal channel with $n=0.5$ are depicted for three different values
of decay time $\gamma{t}$.}
\end{figure}

Recently, the volume $P_{NW}$ of the negative part of Wigner
distribution function has been suggested as a good choice for
quantifying the nonclassicality
\cite{Benedict1999,Dodonov2003,Kenfack2004,Li2007,Li20071}. $P_{NW}$
is defined by \be P_{NW}=|\int_{\Omega}W(q,p)dqdp|,\ee where
$\Omega$ is the negative Wigner distribution region. In
Ref.\cite{Li2007}, we have investigated $P_{NW}$ of photon-added
coherent states in the photon-loss channel. It was shown that
$P_{NW}$ and entanglement potential defined in Ref.\cite{Asboth2005}
exhibit the consistent behaviors in short decay time.

Now, we bring our attention to the influence of thermal noise on the
nonclassicality of the quantum excitation of classical non-coherent
optical thermal fields by calculating $P_{NW}$. Substituting the
time evolving Wigner function in Eq.(8) into Eq.(9), we could obtain
\be
P_{NW}=-\frac{[\frac{\kappa}{2\xi}+2(1+\bar{n})(1-e^{\frac{\kappa}{4(1+\bar{n})\xi}})]e^{-\gamma{t}}e^{\zeta/\xi}}{\xi}
\ee for $\gamma{t}\leq\gamma{t}_c=\ln\frac{2+2n}{1+2n}$. In Fig.2,
$P_{NW}$ of SPATSs with different values of the parameter $\bar{n}$
is plotted as the function of $\gamma{t}$ for two different values
of the mean thermal photon number $n$ of the thermal channel. It is
shown that the thermal noise deteriorates the partial negativity,
and $P_{NW}$ monotonically decreases with the decay time. The more
intense the thermal noise, the more rapidly $P_{NW}$ decreases,
which implies that the nonclassicality of the optical fields are
very fragile against the thermal noise. The above results also
indicate that $P_{NW}$ becomes zero at a threshold decay time
$\gamma{t}_c$ which only depends on the value of $n$ but not
$\bar{n}$. For the case of the SPATSs in thermal channel, the
threshold decay time is \be \gamma{t}_c=\ln(\frac{2+2n}{1+2n}),\ee

\begin{figure}
\centerline{\includegraphics[width=6.3cm]{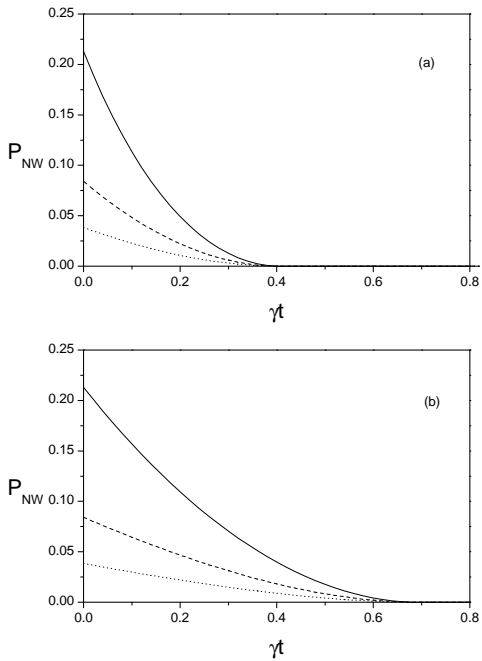}}
\caption{The volume of partial negativity of the evolving Wigner
function of the SPATSs with different values of $\bar{n}$ in the
thermal channels with two different values of mean thermal photon
number (a) $n=0.5$; (b) $n=0$ is plotted as the function of decay
time $\gamma{t}$. (Solid line) $\bar{n}=0$; (Dash line)
$\bar{n}=\frac{3}{7}$; (Dot line) $\bar{n}=1$.}
\end{figure}

Previous studies have indicated that the volume of negative part of
Wigner function of single photon-added coherent states (SPACSs) in
photon-loss channel or thermal channel decreases with the decay
time, and its threshold decay time concerning the complete vanish of
partial negativity of Wigner function does not depend on the seed
intensity \cite{Li2007,Li20071}. In Ref.\cite{Li20071}, we have
presented a relation between the mean photon number $n$ of the
thermal channel and the threshold decay time $\gamma{t}_c(n)$ beyond
which their Wigner function become positive. For arbitrary
nonclassical optical fields which have the partial negative Wigner
distribution function, there exists a relation between the mean
photon number $n$ of the thermal channel and the threshold decay
time $\gamma{t}_c(n)$ \cite{Li20071}, \be
\gamma{t}_c(n)=\ln\frac{e^{\gamma{t}_c(0)}+2n}{1+2n},\ee where
$\gamma{t}_c(0)$ is the threshold decay time in the photon loss
channel.

The similarity between the amplitude-independence of the threshold
decay time $\gamma{t}_c$ corresponding to the disappearance of
partial negativity of the Wigner function of SPACSs and SPATSs in
the thermal channel implies there may exist a universal relation
about $\gamma{t}_c$ for arbitrary single quantum excitation of
classical gaussian states. Here, we demonstrate that all of the
nonclassical states whose density operators $\rho$ satisfy
$\langle0|\rho|0\rangle=0$, where $|0\rangle$ is the vacuum sate,
completely loss the negativity of their Wigner functions at the
threshold decay time $\gamma{t}=\ln\frac{2+2n}{1+2n}$. The strict
proof will be presented elsewhere. Here, we briefly outline the
proof procedures. Firstly, considering the case of photon-loss
channel, i.e. $n=0$. In this situation, based on Eq.(5), for
arbitrary initial nonclassical pure or mixed states whose vacuum
state population is zero, we can derive \be
W(q,p,\gamma{t})|_{\gamma{t}=\ln2}\geq0 \ee for any values of $q$
and $p$, and \be
W(0,0,\gamma{t})|_{\gamma{t}=\ln2}=c\langle0|\rho|0\rangle=0 \ee
where $c$ is a constant. Eqs.(13-14) imply that the threshold decay
time $\gamma{t}_c(0)$ in Eq.(12) is $\ln2$ for these states.
Substituting it into Eq.(12), we can complete this proof. In the
derivation of Eq.(13), we have used the relation between the Wigner
distribution function and the ${\mathcal{Q}}$ function. The
${\mathcal{Q}}$ function gives the probability distribution for
finding the coherent state $|\alpha\rangle$ in the state
$\hat{\rho}$ since
${\mathcal{Q}}(q,p)=\frac{1}{\pi}\langle\alpha|\hat{\rho}|\alpha\rangle$,
where $\alpha=q+ip$. The ${\mathcal{Q}}$ function is always
non-negative. The ${\mathcal{Q}}$ function is a particular case of a
class of non-negative quantum distribution functions, the Husimi
functions, obtained by smoothing the Wigner distribution function
with a minimum uncertainty squeezed Gaussian function \cite{Do}. The
${\mathcal{Q}}$ function can be obtained when the Wigner function is
smoothed by a coherent state wave packet. From Eq.(5), we can find,
for any quantum fields with initial state $\hat{\rho}_0$ in photon
loss channel, their time evolution Wigner functions at decay time
$\gamma{t}=\ln2$ can be rewritten as
$W(q,p,\ln2)\propto\langle\sqrt{2}\alpha|\hat{\rho}_0|\sqrt{2}\alpha\rangle\equiv\pi{\mathcal{Q}}_0(\sqrt{2}q,\sqrt{2}p)$.
Therefore, we have the Eq.(13) by referencing the characteristics of
the ${\mathcal{Q}}$ function.

In summary, we have investigated the nonclassicality of single
photon excitation of thermal optical field in the thermal channel by
exploring the partial negativity of the Wigner function. The total
volume of the negative part defined by the absolute value of the
integral of the Wigner function over the negative distribution
region is analytically calculated. For the case of SPATSs in thermal
channel, the exact threshold value of the decay time beyond which
the evolving Wigner function becomes positive is given as
$\gamma{t}_c=\ln(\frac{2+2n}{1+2n})$, which is the same as the one
in the case of single quantum excitation of the classical coherent
field.

For all of the nonclassical states whose density operators $\rho$
satisfy $\langle0|\rho|0\rangle=0$, where $|0\rangle$ is the vacuum
sate, it is demonstrated that the threshold decay times are the same
and given by $\gamma{t}_c=\ln(\frac{2+2n}{1+2n})$. These results
clearly imply Wigner distributions of any photon-added thermal
states are partial negative before the threshold decay time
$\gamma{t}_c$ even if their initial thermal state seeds are
macroscopic with arbitrary large $\bar{n}$ but finite. Obviously,
any photon-added classical gaussian states including photon-added
thermal states, photon-added coherent states, and photon-added
displaced thermal states belong to the class of states satisfying
$\langle0|\rho|0\rangle=0$. Therefore, the above conclusions can be
generalized to: In thermal channel with mean thermal photon number
$n$, all nonclassical pure or mixed states $\rho$ with zero
population in vacuum state have partial negative Wigner distribution
before the threshold decay time $\gamma{t}_c=\ln\frac{2+2n}{1+2n}$
if ${\mathrm{Tr}}(\rho{a^{\dagger}}a)$ is finite.

The above results can be regarded as benchmark to investigate the
robustness of other indicators of nonclassicality such as squeezing,
antibunch, and entanglement potential of nonclassical optical fields
in thermal channel if compared with the partial negative Wigner
distribution. Recently, the physical realization of controlled phase
gate based on the single-photon-added coherent states has also been
proposed \cite{Li20071}. Our studies in this report may find some
applications in these quantum information processes in which
photon-added states are involved.




\end{document}